\documentclass[12pt]{article}
\usepackage{latexsym}
\usepackage{amsmath}
\usepackage{amssymb}
\catcode`\@=11
\newcount\hour
\newcount\minute
\newtoks\amorpm \hour=\time\divide\hour by 60\minute
=\time{\multiply\hour by 60 \global\advance\minute by-\hour}
\edef\standardtime{{\ifnum\hour<12 \global\amorpm={am}%
        \else\global\amorpm={pm}\advance\hour by-12 \fi
        \ifnum\hour=0 \hour=12 \fi
        \number\hour:\ifnum\minute<10
        0\fi\number\minute\the\amorpm}}
\edef\militarytime{\number\hour:\ifnum\minute<10
0\fi\number\minute}
\def\draftlabel#1{{\@bsphack\if@filesw {\let\thepage\relax
   \xdef\@gtempa{\write\@auxout{\string
      \newlabel{#1}{{\@currentlabel}{\thepage}}}}}\@gtempa
   \if@nobreak \ifvmode\nobreak\fi\fi\fi\@esphack}
        \gdef\@eqnlabel{#1}}
\def\@eqnlabel{}
\def\@vacuum{}
\def\marginnote#1{}
\def\draftmarginnote#1{\marginpar{\raggedright\scriptsize\tt#1}}
\def\draft{
        \pagestyle{plain}
        \overfullrule=2pt
        \oddsidemargin -.1truein
        \def\@oddhead{\sl \phantom{\today\quad\militarytime} \hfil
        \smash{\Large\sl DRAFT} \hfil \today\quad\militarytime}
        \let\@evenhead\@oddhead
        \let\label=\draftlabel
        \let\marginnote=\draftmarginnote
        \def\ps@empty{\let\@mkboth\@gobbletwo
        \def\@oddfoot{\hfil \smash{\Large\sl DRAFT} \hfil}
        \let\@evenfoot\@oddhead}
        \def\@eqnnum{(\theequation)\rlap{\kern\marginparsep\tt\@eqnlabel}%
        \global\let\@eqnlabel\@vacuum}  }

\renewcommand{\theequation}{\thesection.\arabic{equation}}
\renewcommand{\thefootnote}{\fnsymbol{footnote}}

\def\appendix#1{\addtocounter{section}{1}\setcounter{equation}{0}
\renewcommand{\thesection}{\Alph{section}}
\section*{Appendix \thesection\protect\indent \parbox[t]{11.15cm}{#1}}
\addcontentsline{toc}{section}{Appendix \thesection\ \ \ #1}}

\def \la {\label}

\def\be{\begin{equation}}
\def\ee{\end{equation}}

\catcode`\@=11

\def\bea{\begin{eqnarray}}
\def\eea{\end{eqnarray}}
\def\beann{\begin{eqnarray*}}
\def\eeann{\end{eqnarray*}}
\def\beq{\begin{equation}}
\def\eeq{\end{equation}}
\def\ba{\begin{array}}
\def\ea{\end{array}}
\def\ben{\begin{enumerate}}
\def\een{\end{enumerate}}

 \def \la {\label}
 \def\be{\begin{equation}}
\def\ee{\end{equation}}

\def \la {\label}

\font\mybb=msbm10 at 11pt

\def\bb#1{\hbox{\mybb#1}}

\def\bR {\bb{R}}

\def\bC {\bb{C}}

\def \ee {\epsilon}

\def\be{\begin{equation}}
\def\ee{\end{equation}}

\def \la{\label}

\begin{document}
\date{November 2002}
\begin{titlepage}
\begin{center}
\hfill KUL-TF-06/29 \\
\hfill UB-ECM-PF-06-35 \\
\hfill hep-th/0611150 \\

\vspace{2.0cm} {\Large \bf Classification of supersymmetric backgrounds \\ of
  string theory} \\[.2cm]

\vspace{1.5cm}
 {\large  U. Gran$^1$, J. Gutowski$^2$,  G. Papadopoulos$^3$ and D. Roest$^4$}

\vspace{0.5cm}

${}^1$ Institute for Theoretical Physics, K.U. Leuven\\
Celestijnenlaan 200D\\
B-3001 Leuven, Belgium\\

\vspace{0.5cm}
${}^2$ DAMTP, Centre for Mathematical Sciences\\
University of Cambridge\\
Wilberforce Road, Cambridge, CB3 0WA, UK

\vspace{0.5cm}
${}^3$ Department of Mathematics\\
King's College London\\
Strand\\
London WC2R 2LS, UK\\

\vspace{0.5cm}
${}^4$ Departament Estructura i Constituents de la Materia \\
    Facultat de F\'{i}sica, Universitat de Barcelona \\
    Diagonal, 647, 08028 Barcelona, Spain \\

\end{center}

\vskip 1.5 cm
\begin{abstract}

We review the recent progress made towards the classification of supersymmetric solutions in ten and eleven dimensions with emphasis
  on those of IIB supergravity. In particular, the spinorial geometry method is outlined and adapted to nearly maximally
  supersymmetric backgrounds.
We then demonstrate its effectiveness  by classifying the maximally supersymmetric IIB
$G$-backgrounds and by showing that $N=31$ IIB solutions do not exist.

\end{abstract}
\vskip 0.5cm

\line(1,0){50}

{\small {Based on a talk given by D.R.~on the RTN project `Constituents,
  Fundamental Forces and Symmetries of the Universe'
  conference in Napoli, October 9 - 13, 2006. }}
\end{titlepage}
\newpage
\setcounter{page}{1}
\renewcommand{\thefootnote}{\arabic{footnote}}
\setcounter{footnote}{0}

\setcounter{section}{0}
\setcounter{subsection}{0}

\section{Introduction}

The supersymmetric solutions of $D=10$ and $D=11$ supergravities are
instrumental in the understanding of string/M-theory dualities,
compactifications and the AdS/CFT correspondence. Most of these solutions
have been found using Ans\"atze adapted to the requirements of
physical problems. However, the realization that there are new
maximally supersymmetric solutions \cite{georgea}, the
rediscovery of some old ones \cite{kowalski, georgeb}, and their subsequent applications in AdS/CFT correspondence,
  have indicated that a
more systematic investigation  of supersymmetric solutions in
supergravity theories is needed. By solving ${\cal R}=0$, where ${\cal R}$ is the
supercovariant curvature,
 the authors of \cite{jffgp} classified the    maximally supersymmetric solutions of  $D=10$
and $D=11$ supergravities.  The $G$-structure method,  based on the Killing spinor form  bi-linears and refined in \cite{oisin},  has
also been used in \cite{pakis, gutowski} to solve the Killing spinor equations (KSE) of
$N=1$ backgrounds of $D=11$ supergravity, i.e. the backgrounds that admit one Killing spinor.

The spinorial geometry method of solving Killing spinor equations, proposed in \cite{joe}, is based on
the gauge symmetry of
Killing spinor equations, on  a
description of spinors in terms of forms,  and on an oscillator basis in the space of
spinors. In $D=11$, it has been applied to considerably simplify the solution of the KSE
for $N=1$ backgrounds, and then to investigate backgrounds with   two, three,  four and 31 supersymmetries. Furthermore, spinorial geometry
has been
used to solve the KSE of IIB $N=1$  backgrounds \cite{ugjggpa, ugjggpb}, and
to explore the geometry
of supersymmetric heterotic backgrounds \cite{het}. In this talk
two of the most recent applications in IIB are reviewed. These are the supersymmetric
backgrounds with the maximal number of $G$-invariant Killing spinors
\cite{ugjggpb, IIBmax}, and $N=31$ backgrounds
\cite{IIBpreons}.

\section{IIB maximally supersymmetric $G$-backgrounds}

Supersymmetric backgrounds can be
characterized by the number $N$ of Killing spinors  and their
stability subgroup $G$ in an appropriate spin group \cite{josew}.
For a given stability subgroup $G$,  the KSE of IIB
supergravity simplify for backgrounds that admit the maximal number
of $G$-invariant Killing spinors \cite{ugjggpb, IIBsystem}. Such backgrounds can be thought of as
 the vacua of IIB strings in a compactification senario. In particular, it has been found that the Killing
spinors can be written as
 \begin{eqnarray}
 \epsilon_i=\sum_{j=1}^N\,f_{ij}\,\eta_j~, \qquad j=1, \dots, N=2m~,
 \end{eqnarray}
where $\eta_p$, $p=1, \dots, m$ is a basis of $G$-invariant
Majorana-Weyl spinors, $\eta_{m+p}=i\eta_p$, and $(f_{ij})$ is an
$N\times N$ invertible matrix of real spacetime functions. It turns out that in such cases
the IIB KSE and their integrability
conditions factorize \cite{ugjggpb, IIBsystem}.

 IIB Killing spinors are invariant under the   subgroups
$Spin(7)\ltimes\mathbb{R}^8~ (2)$, $SU(4)\ltimes\mathbb{R}^8~(4)$,
$Sp(2)\ltimes\mathbb{R}^8$ (6), $(SU(2)\times SU(2))\ltimes\mathbb{R}^8~(8)$,
$\mathbb{R}^8~ (16)$, $G_2~(4)$, $SU(3)~(8)$, $SU(2)~(16)$ and
$\{1\}~(32)$ of $Spin(9,1)$, where the number in parenthesis denotes the maximal number of invariant spinors
 in each case. These groups have been found in the context of the heterotic string \cite{het}. The $\{1\}~(N=32)$ case
 consists of the maximally supersymmetric backgrounds which have been classified in \cite{jffgp}. These are locally isometric
 to Minkowski spacetime
$\mathbb{R}^{9,1}$, $AdS_5\times S^5$ \cite{schwarz} and the maximally
supersymmetric Hpp-wave \cite{georgea}. The remaining cases have been classified in
\cite{ugjggpb, IIBmax}. It is instructive to distinguish between compact and non-compact stability subgroups in  $Spin(9,1)$
because the geometry is different in these two cases.

{}First consider the the supersymmetric backgrounds associated with the compact stability subgroups $G=G_2, SU(3)$ and $SU(2)$.
The
spacetime $M$ of such backgrounds  is locally isometric to a product $M=X_n\times Y_{10-n}$
with $n=3,4,6$, where $X_n$ is a maximally supersymmetric solution
of an $n$-dimensional supergravity theory and $Y_{10-n}$ is a
Riemannian manifold with holonomy $G$. In the $G_2$ case $X_3=\bR^{2,1}$ and $Y_7$ is a holonomy $G_2$ manifold. In the $SU(3)$ case,
$X_4=AdS_2\times S^2$, $\bR^{3,1}$ or $CW_4$ and $Y_6$ is a Calabi-Yau manifold, where $CW_4$ is a 4-dimensional Cahen-Wallach plane wave.
Similarly in the $SU(4)$ case, $X_6=AdS_3\times S^3$, $\bR^{5,1}$ or $CW_6$
and $Y_4$ is hyper-K\"ahler. Apart from the cases in which $X_n$
is flat, all these backgrounds have non-trivial fluxes and the full solutions can be found in  \cite{IIBmax}.

Next we summarize the geometry and fluxes of supersymmetric backgrounds associated with non-compact
stability subgroups $G=K\ltimes\mathbb{R}^8$ for
$K=Spin(7), SU(4), Sp(2)$, $SU(2)\times SU(2)$ and $\{1\}$, for a detailed exposition see \cite{IIBmax}. In all these cases,
 the spacetime $M$  admits a null parallel vector field $X$ and the {\it holonomy of the  Levi-Civita
connection}, $\nabla$, of spacetime is contained in $K\ltimes\mathbb{R}^8$, i.e.
\begin{eqnarray}
\nabla_A X=0~,~~~~{\rm hol}(\nabla)\subseteq K\ltimes\mathbb{R}^8~.
\end{eqnarray}
Therefore, the spacetime
is a pp-wave propagating in an eight-dimensional Riemannian
 manifold
$Y_8$ such that ${\rm hol}(\tilde \nabla)\subseteq K$, where $\tilde \nabla$ is the Levi-Civita connection of $Y_8$.
Alternatively, the spacetime is a two-parameter Lorentzian  deformation family of $Y_8$. Adapting
coordinates along the parallel vector field $X=\partial/\partial u$, the metric can be written as
\begin{eqnarray}
ds^2=2 dv (du+Vdv+n)+ds^2(Y_8)~,~~~~ds^2(Y_8)=\gamma_{IJ}dy^I dy^J=\delta_{ij} e^i_I e^j_J dy^I dy^J
\la{metrnull}
\end{eqnarray}
where $V$, $n$, and the metric $ds^2(Y_8)$ may also depend on the coordinate $v$.
In all cases, the fluxes are null,
\begin{eqnarray}
P=P_-(v) e^-~,~~~G=e^-\wedge L~,~~~F=e^-\wedge M~, \la{fluxLM}
\end{eqnarray}
and the
Bianchi identities give $dP=dG=dF=0$, where $L$ and $M$ are a two-
and a self-dual  four-form, respectively, of $Y_8$. In particular,
one finds that $P_-=P_-(v)$. Let $\mathfrak{k}$ be the Lie algebra of $K$. To give the
conditions that the Killing spinor equations impose on the fluxes,
 decompose $L\in \Lambda^2(\mathbb{R}^8)\otimes \mathbb{C}$  and $M\in
\Lambda^{4+} (\mathbb{R}^8)$ in irreducible representations of $K$ as
\begin{eqnarray}
L=L^{\mathfrak{k}}+L^{\rm
inv}~,~~~M= M^{\rm inv}+\tilde M~, \la{fluxLMinv}
\end{eqnarray}
where
$L^{\mathfrak{k}}$ is the Lie algebra valued component of $L$
 in the decomposition $\Lambda^2(\mathbb{R}^8)=\mathfrak{k}\oplus\mathfrak{k}^\perp$,
and $L^{\rm inv}$ and $M^{\rm inv}$ are $K$-invariant two- and
four-forms, respectively. $M^{\rm inv}$ decomposes further as
$M^{\rm inv}=m^0+\hat M^{\rm inv}$, where $m^0$ has the property
that the associated Clifford algebra element satisfies
$m^0\epsilon=g \epsilon$, $g\not=0$ a spacetime function, for all
Killing spinors $\epsilon$. In a particular gauge, the Killing
spinor equations imply that $g$ is proportional to $Q_-$
 and  restrict   the spacetime dependence
of  $L^{\rm inv}$ and
 $M^{\rm inv}$.
Furthermore,  $\tilde M$ takes values in a representation of $K$ in
$\Lambda^{4+} (\mathbb{R}^8)$ with the property that the associated
Clifford algebra element satisfies $\tilde M\epsilon=0$ for all
Killing spinors $\epsilon$. $L^{\mathfrak{k}}$ and $\tilde M$ are
not restricted by the Killing spinor equations.

{}For compact stability subgroup $G$, the Killing spinor equations imply  the field equations provided
the Bianchi identities are satisfied.  For
the non-compact $G$, the integrability conditions of the Killing spinor equations and the Bianchi identities imply
that all field equations are satisfied provided that $E_{--}=0$, where $E_{--}$ denotes the
$--$ component of the Einstein equations. This in turn gives
\begin{eqnarray}
  && - (\partial^i + \Omega_{j,}{}^{ji})(\partial_i V -
  \partial_v n_I e^I{}_i) + \tfrac{1}{4} (dn)_{ij} (dn)^{ij} -
  \tfrac{1}{2} \gamma^{IJ} \partial_v{}^2 \gamma_{IJ}
  - \tfrac{1}{4} \partial_v \gamma^{IJ} \partial_v \gamma_{IJ} \nonumber \\
  &&
 - \tfrac{1}{6} F_{- i_1 \cdots i_4} F_-{}^{i_1 \cdots i_4}
  - \tfrac{1}{4} G_{-}{}^{i_1 i_2} G^*_{- i_1 i_2} -2 P_- P_-^* = 0 \,,
 \label{Einstein-eq}
 \end{eqnarray}
where $\gamma^{IJ}$ is the inverse of the metric $\gamma_{IJ}$
defined in ({\ref{metrnull}}). For the special
case of  fields  independent of $v$, this equation
becomes
 \begin{eqnarray}
  - \Box_8 V + \tfrac{1}{4} (dn)_{ij} (dn)^{ij} - \tfrac{1}{6} F_{- i_1 \cdots i_4} F_-{}^{i_1 \cdots i_4}
  - \tfrac{1}{4} G_{-}{}^{i_1 i_2} G^*_{- i_1 i_2} -2 P_- P_-^* = 0 \,,
 \label{Einstein-eq-trunc}
 \end{eqnarray}
where $\Box_8$ is the Laplacian on the eight-dimensional space $Y_8$ and
$dn$  takes values in $\mathfrak{k}$. Observe that the spacetime rotation and the fluxes contribute with different relative
sign in the field equation as may have been expected.

The backgrounds that we have found can be thought of as vacua of IIB
string theory. This particulary applies to compact stability
subgroups. The backgrounds  $\mathbb{R}^{9-n, 1}\times Y_n$ are
vacua of IIB compactifications on $G_2$ for $n=7$, and on Calabi-Yau
manifolds for $n=6$ and $n=4$. The backgrounds $AdS_{5-n/2}\times
S^{5-n/2}\times Y_n$ can be thought of as either the vacua of the
Calabi-Yau or $S^{5- n/2}\times Y_n$ compactifications with fluxes.
It is worth pointing out that there are additional vacua associated with the plane waves.

\section{N=31 in IIB}

The holonomy of the
supercovariant connection of type II and $D=11$ supersymmetric, $N<32$,  backgrounds  is a proper subgroup of
$SL(32,\mathbb{R})$. In particular  for any $N<32$, there are components of the supercovariant curvature
which are not restricted by the gravitino KSE,  for M-theory see
\cite{hull, duffl, gpta}  and for IIB see \cite{gptb}.
Furthermore, it was argued in \cite{IIBsystem} that the Killing
spinor bundle ${\cal K}$ can be any subbundle of the Spin bundle and
the spacetime geometry depends on the trivialization of ${\cal K}$.
This is unlike what happens in the case of Riemannian and Lorentzian
geometries \cite{berger, josew} and heterotic and type I
supergravities (provided the parallel spinors are
Killing) \cite{heterotic} where there are restrictions both on the
number of Killing spinors and the Killing spinor bundle. It is clear from the above arguments that the gravitino KSE allows for the possibility
that   supersymmetric backgrounds exist
for any $N$. However, the algebraic KSE, Bianchi identities and field equations that supersymmetric backgrounds
must satisfy are not included in the holonomy argument. Because of this, the holonomy argument is not conclusive.

To investigate whether there are backgrounds for any $N$ we consider IIB $N=31$ supersymmetric backgrounds.
Backgrounds with 31 supersymmetries have
been considered  in the context of M-theory \cite{bandos} and
have been called preons.
We shall see that the IIB algebraic KSE implies that such backgrounds must be maximally supersymmetric \cite{IIBpreons}.
  To our knowledge this is the first
example which demonstrates that there are restrictions on the number
of supersymmetries of type II backgrounds. To do this, we shall
adapt the spinorial method \cite{joe} of solving Killing spinor
equations to backgrounds that admit near maximal number of
supersymmetries. We shall mostly focus on IIB
but the analysis extends to $D=11$ and other lower-dimensional supergravities.

To adapt the spinorial method to backgrounds with near maximal
number of supersymmetries, we
 introduce   a ``normal'' ${\cal K}^\perp$ to the Killing spinor bundle ${\cal K}$ of a supersymmetric background.
 The spinors of IIB supergravity
are complex positive chirality Weyl spinors, so the Spin bundle is
${\cal S}^c_+={\cal S}_+\otimes\mathbb{C}$, where ${\cal S}_+$ is the rank
sixteen  bundle of positive chirality Majorana-Weyl spinors. ${\cal
S}^c_+$  may also be thought of as an associated bundle of a
principal bundle with fibre $SL(32, \mathbb{R})$, the holonomy group of the
supercovariant connection, acting with the fundamental
representation on $\mathbb{R}^{32}$. If a background admits $N$ Killing
spinors which span the fibre of the Killing spinor bundle ${\cal
K}$, then one has the sequence
\begin{eqnarray} 0\rightarrow {\cal K}\rightarrow
{\cal S}_+^c\rightarrow {\cal S}_+^c/{\cal K}\rightarrow 0~.
\la{exseq}
\end{eqnarray}
The inclusion   $i:\,{\cal K}\rightarrow {\cal
S}_+^c$ can be locally described as \begin{eqnarray} \epsilon_r=\sum_{i=1}^{32}
f^i{}_r \eta_i~,~~~r=1,\dots, N \end{eqnarray} where $\eta_p$, $p=1,\dots,16$,
is a basis in the space of positive chirality Majorana-Weyl spinors,
$\eta_{16+p}= i\eta_p$ and the coefficients $f$ are real spacetime
functions. For our notation and spinor conventions see
\cite{IIBsystem}. Any $N$  Killing spinors related by a local
$Spin(9,1)$ transformation give rise to the same spacetime geometry.
This is because the Killing spinor equations and the field equations
of IIB supergravity are Lorentz invariant. Therefore any bundles of
Killing spinors and any choice of sections related by a $Spin(9,1)$
gauge transformation should be identified.

The normal to the Killing spinor bundle, ${\cal K}^\perp$, is a subbundle of ${\cal S}_-^c=S_-\otimes\bC$
defined  by the orthogonality condition
\begin{eqnarray}
{\cal B}(\nu,\epsilon)=0~,
\la{normcol}
\end{eqnarray}
where $\nu$ is a section of ${\cal K}^\perp$, ${\cal B}={\rm Re}\, B$ is a non-degenerate inner product, and $B: {\cal S}_+\otimes
{\cal S}_-\rightarrow \mathbb{R}$ is a $Spin(9,1)$-invariant inner product
\begin{eqnarray}
B(\epsilon, \zeta)=-B(\zeta,
\epsilon)=<B(\epsilon^*), \zeta>~,
\end{eqnarray}
extended bi-linearly on ${\cal S}^c_+ \otimes {\cal S}^c_-$.
 To write this orthogonality condition in components,
introduce  a basis in ${\cal S}^c_-$, say
$\theta_{i'}=-\Gamma_0\eta_i$. Then write $\nu=n^{i'}
\theta_{i'}$ and the condition (\ref{normcol}) can be written as
\begin{eqnarray}
n^{i'} {\cal B}_{i' j} f^j{}_r=0~,
\la{normcolcom}
\end{eqnarray}
where
${\cal B}_{i'j}={\cal B}(\theta_{i'}, \eta_j)$.

Let us now consider the IIB $N=31$ backgrounds. The rank of ${\cal K}^\perp$ is one.
The spinorial geometry method can be applied as follows. First the $Spin(9,1)$ gauge symmetry
of the IIB KSE can be used to orient the normal spinor along three different directions.
This is because there are three kinds of orbits of $Spin(9,1)$
in the negative chirality Weyl spinors with stability subgroups
$Spin(7)\ltimes\mathbb{R}^8$, $SU(4)\ltimes \mathbb{R}^8$ and $G_2$, respectively. This can be easily seen
using the results of  \cite{ugjggpa}.
The three representatives can be chosen as
 \begin{eqnarray}
  &&\nu_1=(n+im)
(e_5+e_{12345})~,~~~\nu_2= (n-\ell+im) e_5+ (n+\ell+im) e_{12345}~,
\cr
&&\nu_3=n(e_5+e_{12345})+i m (e_1+e_{234})~,
\end{eqnarray}
where according to spinorial geometry we have written the spinors as multi-forms.
Therefore up to a  $Spin(9,1)$ gauge transformation,  ${\cal K}^\perp$ can be chosen to lie along one of these three
directions. In turn enforcing the orthogonality condition (\ref{normcolcom}), there are three different hyper-planes
that the Killing spinors lie in the space of spinors. The expressions for the Killing spinors can be found in \cite{IIBpreons}.
Next, one substitutes  the Killing spinors into the IIB algebraic KSE. Then either a direct
computation using an oscillator basis in the space of spinors or a straightforward argument
based on the expression of Killing spinors in terms of forms reveals that the
flux field strengths $P$ and $G$ vanish, $P=G=0$. Due to this, the IIB gravitino Killing spinor equation becomes
linear over the complex numbers. This means that backgrounds with
vanishing $P$ and $G$ fluxes always preserve an even number of
supersymmetries. Thus backgrounds with 31 supersymmetries preserve
an additional supersymmetry and so they are maximally
supersymmetric.  Later it was shown that IIA $N=31$ backgrounds are also maximally supersymmetric in \cite{bandosb}.
Thus there are no type II preons.

Next let us consider $N=31$ backgrounds in eleven dimensions. $D=11$ supergravity does not have an
algebraic KSE and so the analysis presented for such backgrounds in type II theories does not generalize.
Nevertheless, the spinorial geometry method
can be easily adapted to investigate $N=31$ supersymmetric backgrounds in eleven dimensions. In particular,
one can show that the Killing spinors, after an appropriate choice of the normal spinor
up to $Spin(10,1)$ gauge transformations, take a rather simple form. Next,  the holonomy argument indicates that there may be
$D=11$ backgrounds with $N=31$ supersymmetries. But it turns out that all components of the supercovariant curvature vanish
as a consequence of imposing in addition the Bianchi identities and the field equations of the theory. Thus the reduced
holonomy of $N=31$ backgrounds is in fact $\{1\}$ and so these backgrounds admit  an additional Killing spinor. Therefore the $N=31$
backgrounds are locally isometric to maximally supersymmetric ones \cite{Mpreons}. Similar
results also hold for type I supergravity. Therefore in $D=10$ and $D=11$ supergravities there are not supersymmetric
backgrounds for all $N$. This may lead to a simplification in the classification
of supersymmetric backgrounds.

\section*{Acknowledgement}
  
D.R.~wishes to thank the organisers of the RTN project `Constituents,
  Fundamental Forces and Symmetries of the Universe'
  conference in Napoli for the opportunity to give a talk in a very nice and
  inspiring atmosphere.
Part of the work on which this talk is based was completed while D.R.~was a post-doc at King's College
London, for which he would like to acknowledge the PPARC grant
PPA/G/O/2002/00475. In addition, he is presently supported by the European
EC-RTN project MRTN-CT-2004-005104, MCYT FPA 2004-04582-C02-01 and CIRIT GC
2005SGR-00564. U.G.~has a
postdoctoral fellowship funded by the Research Foundation
K.U.~Leuven.

\end{document}